\documentclass[aps,prc,superscriptaddress,reprint,floatfix]{revtex4-2}
\usepackage{graphicx}
\usepackage{multirow}
\usepackage{color}
\usepackage{amssymb}
\usepackage{bbm}
\usepackage{amsmath}
\usepackage{lineno}
\usepackage{enumitem}
\usepackage{scalerel,stackengine}
\usepackage{xcolor}
\usepackage{comment}
\usepackage{booktabs}
\usepackage[normalem]{ulem}
\usepackage[colorlinks=true, breaklinks=true, linkcolor=blue, citecolor=purple, urlcolor=teal]{hyperref}
\usepackage{algorithm}
\usepackage{algpseudocode}

\renewcommand{\Im}{\mathop{\mathrm{Im}}}
\renewcommand{\Re}{\mathop{\mathrm{Re}}}

\def\nuc#1#2{\relax\ifmmode{}^{#1}{\protect\text{#2}}\else${}^{#1}$#2\fi}
\newcommand{\etal}{\textit{et al.~}}

\newcommand{\be}{\begin{eqnarray}}
\newcommand{\ee}{\end{eqnarray}}

\newcommand{\bwt}{\begin{widetext}}
\newcommand{\ewt}{\end{widetext}}

\begin{document}

\title{Bidirectional Neural Networks for Global Nucleon-Nucleus Optical Model Calculations}

\author{Jin Lei}
\email[]{jinl@tongji.edu.cn}
\affiliation{School of Physics Science and Engineering, Tongji University, Shanghai 200092, China.}
\affiliation{Southern Center for Nuclear-Science Theory (SCNT), Institute of Modern Physics, Chinese Academy of Sciences, Huizhou 516000, Guangdong Province, China.}

\begin{abstract}
Modern nuclear data evaluation increasingly requires not only accurate scattering calculations, but also efficient methods for uncertainty quantification and parameter optimization, tasks that benefit from differentiable solvers amenable to gradient-based algorithms. I present a neural network emulator based on Bidirectional Liquid Neural Networks (BiLNN) that provides a differentiable mapping from the optical potential to scattering wave functions. The key innovation enabling generalization across the parameter space is the use of phase-space coordinates $\rho = kr$ that normalize the oscillation wavelength regardless of projectile energy, allowing a single network to span 1 to 200~MeV. Trained on Numerov solutions for twelve target nuclei (\nuc{12}{C} to \nuc{208}{Pb}), both protons and neutrons, and partial waves up to $l=30$, the network achieves an overall relative error of 0.6\%. The predicted wave functions yield accurate $S$-matrix elements and elastic scattering cross sections, reproducing diffraction patterns spanning four orders of magnitude. Importantly, the model generalizes successfully to nuclei not included in training (\nuc{24}{Mg}, \nuc{63}{Cu}, \nuc{184}{W}) with comparable accuracy, demonstrating that it has captured the smooth functional dependence encoded in the KD02 parameterization rather than memorizing specific targets. The differentiable nature of the trained model is the key prerequisite for gradient-based optimization of optical model parameters and uncertainty quantification, which I identify as the principal direction for future work.
\end{abstract}


\date{\today}

\maketitle

\section{Introduction}
\label{sec:intro}

Nuclear reactions are central to many areas of physics, from understanding the synthesis of elements in stars to applications in nuclear energy and medicine. When a nucleon, a proton or neutron, collides with an atomic nucleus, a variety of processes can occur: the nucleon may scatter elastically, preserving the internal state of both projectile and target; it may excite the target nucleus to higher energy levels; or it may be absorbed, leading to compound nucleus formation or other non-elastic channels. The relative probabilities of these outcomes, encoded in scattering cross sections, depend sensitively on the interaction between the incident nucleon and the target nucleus. Accurate theoretical predictions of these cross sections require a reliable description of the nucleon-nucleus potential and efficient methods for solving the resulting quantum mechanical scattering problem.

The optical model provides the standard theoretical framework for nucleon-nucleus scattering~\cite{Hodgson1971,Satchler1983}. The complex many-body interaction is replaced by an effective one-body potential with real and imaginary parts: the real part describes elastic scattering, while the imaginary part accounts for absorption into non-elastic channels. The potential is typically parameterized using Woods-Saxon functions with 10--20 adjustable parameters. Among the most widely used parameterizations is the Koning-Delaroche (KD02) potential~\cite{Koning2003}, which describes both proton and neutron scattering on nuclei from magnesium to bismuth at energies from 1~keV to 200~MeV.

Given the optical potential, the scattering observables are obtained by solving the radial Schr\"odinger equation for each partial wave. The radial wave function must vanish at the origin and match to the appropriate asymptotic form, either Coulomb wave functions for charged particles or spherical Bessel functions for neutrons, at large distances where the nuclear potential vanishes. From this asymptotic matching, the scattering matrix element is extracted, and the elastic scattering angular distribution and reaction cross section follow from standard partial-wave expressions~\cite{Thompson2009}.

The standard numerical approach for solving the radial Schr\"odinger equation is the Numerov algorithm, a fourth-order finite-difference method that is highly accurate and stable. However, the Numerov algorithm, like all traditional numerical integrators, is fundamentally a discrete, step-by-step procedure that cannot be naturally embedded into modern gradient-based optimization frameworks. While numerical derivatives (finite differences) can be and have been used for gradient-based optimization of optical potential parameters, most notably within the sfresco framework~\cite{Thompson1988,Thompson2009}, the cost of numerical differentiation scales linearly with the number of parameters, and accuracy degrades in regions of parameter space where observables change rapidly. Automatic differentiation, in contrast, provides exact gradients at a cost comparable to a single forward evaluation, regardless of parameter dimensionality. This limitation becomes significant as nuclear physics increasingly adopts inverse problem methodologies, where one seeks to infer potential parameters from experimental observables.

The limitations of traditional numerical methods become acute in the context of modern nuclear physics research. Next-generation radioactive beam facilities such as FRIB and FAIR are producing data on nuclei far from stability, requiring optical model predictions for systems where experimental constraints are sparse~\cite{Aumann2013,Hebborn_2023}. Nuclear astrophysics demands reaction rates for thousands of nuclei across wide energy ranges to model stellar nucleosynthesis~\cite{Arcones2023}. Meanwhile, rigorous uncertainty quantification has become essential in nuclear reaction theory, requiring systematic exploration of high-dimensional parameter spaces through both frequentist methods such as parametric bootstrap with $\chi^2$ confidence bands~\cite{Lovell2017,King2018} and Bayesian inference~\cite{King2019,CatacoraRios2019,Lovell2021,Catacora2021}. In the related field of nuclear data evaluation, Bayesian Monte-Carlo methods have also been adopted for systematic uncertainty propagation~\cite{Koning2015}. Recent developments in uncertainty-quantified global optical potentials, including the KDUQ and CHUQ potentials by Pruitt \etal~\cite{Pruitt2023} and the WLM potential by Whitehead, Lim, and Holt~\cite{Whitehead2021}, underscore the need for efficient emulators that can evaluate millions of parameter samples. Reduced-order emulators meet this need by projecting the high-fidelity solver onto a small subspace spanned by precomputed solutions: eigenvector continuation~\cite{Frame2018,Furnstahl2020}, the reduced-basis method~\cite{Drischler2023}, and their applications to scattering and coupled-channels reactions, including ROSE~\cite{Odell2024}, wave-function-based coupled-channels emulation~\cite{CatacoraRios2026}, a complex-scaling-based scattering emulator~\cite{LiuJunzhe2024}, and a continuum-discretized coupled-channels emulator~\cite{LeiRBM2026}, all reach high accuracy (often at or below the 0.1\% level) but construct a problem-specific basis that must be rebuilt for each new system. The present work takes a complementary, non-intrusive route: rather than projecting onto per-problem snapshots, a single neural network is trained once and then generalizes across nuclei, energies, and partial waves, trading some accuracy for breadth of applicability. These developments call for solvers that are not merely fast, but \emph{differentiable}, enabling gradients of observables with respect to potential parameters to be computed efficiently through automatic differentiation.

Machine learning offers precisely this capability. A neural network that maps potential parameters to wave functions is inherently differentiable: gradients flow seamlessly from the output wave function back through the network to the input parameters via the chain rule. This enables end-to-end gradient propagation through the surrogate, which can complement existing optimization and Bayesian workflows by providing exact gradients at a cost comparable to a single forward evaluation, reducing reliance on repeated numerical differentiation. The neural network thus serves not as a faster replacement for Numerov, but as a qualitatively different computational object, a \emph{differentiable surrogate} that bridges the gap between forward modeling and inverse inference. Beyond differentiability, neural networks can discover and exploit regularities in the solution space. If the optical model captures genuine physical universality, as suggested by the success of global parameterizations, then a well-designed network should learn this universality and generalize to new nuclei not seen during training.

Neural networks have been applied successfully to various problems in nuclear physics, including predictions of nuclear masses~\cite{Niu2018,Lovell2022,Mumpower2022} and level densities~\cite{Lasseri2020}. Physics-informed approaches have emerged as powerful tools for solving differential equations~\cite{Raissi2019,Karniadakis2021}, and specialized architectures have been developed for quantum mechanical problems~\cite{Mills2017,Hermann2020}. However, applying these methods to scattering wave functions presents unique challenges.

A significant challenge in applying neural networks to the scattering problem is the wide range of length scales involved. The de Broglie wavelength of the projectile varies from approximately 9~fm at 10~MeV to about 2~fm at 200~MeV, a factor of 4.5 variation across this 10--200~MeV range. A network trained to predict wave functions as a function of the radial coordinate would need to learn oscillatory patterns with vastly different periods at different energies, which, while not orders of magnitude in span, still presents a meaningful challenge for a single network architecture, as observed empirically in my preliminary experiments. Moreover, the wave function amplitude can vary over several orders of magnitude due to absorption effects from the imaginary potential, adding another layer of complexity.

In this work, I address these challenges by developing a neural network architecture specifically designed for the nucleon-nucleus scattering problem. Two key innovations enable generalization across the full parameter space of energies, partial waves, and target nuclei. First, I introduce \emph{phase-space coordinates} that absorb the energy dependence of the wavelength: in this representation, the wave function oscillates with a universal period regardless of energy, dramatically simplifying the learning task. Second, I employ a \emph{Bidirectional Liquid Neural Network} (BiLNN) based on Closed-form Continuous-time (CfC) layers~\cite{Hasani2022}, which process the radial coordinate sequence in both forward and backward directions. This bidirectional architecture naturally incorporates the two boundary conditions that define the scattering problem: the wave function vanishing at the origin and the asymptotic Coulomb behavior at large distances.

The network is trained on wave functions computed using the Numerov method with the KD02 optical potential. The training set spans twelve target nuclei from \nuc{12}{C} to \nuc{208}{Pb}, both proton and neutron projectiles, energies randomly sampled from a uniform distribution over 1 to 200~MeV, and partial waves from 0 to 30. This comprehensive coverage ensures that the trained network can interpolate reliably across the full parameter space relevant for practical applications. The network reproduces the Numerov wave functions with an average relative error of 0.6\% across the training domain. The trained model provides a fast, differentiable surrogate; exploiting that differentiability for Bayesian inference and sensitivity analysis is left to future work.

The paper is organized as follows. Section~\ref{sec:theory} presents the theoretical framework for nucleon-nucleus scattering, including the optical model potential and the Schr\"odinger equation. Section~\ref{sec:network} describes the neural network architecture, the phase-space coordinate representation, and the training procedure. Results for wave function predictions and scattering observables are presented in Sec.~\ref{sec:results}. Finally, conclusions and an outlook for future developments are given in Sec.~\ref{sec:conclusions}.

\section{Theoretical Framework}
\label{sec:theory}

The scattering of a nucleon from a nucleus is described by the radial Schr\"odinger equation
\begin{equation}
\frac{d^2 u_l}{dr^2} + \left[ k^2 - U(r) - \frac{l(l+1)}{r^2} \right] u_l(r) = 0,
\label{eq:schrodinger}
\end{equation}
where $u_l(r)$ is the radial wave function for partial wave $l$, $k = \sqrt{2\mu E_{\text{cm}}}/\hbar$ is the asymptotic wave number with reduced mass $\mu$ and center-of-mass energy $E_{\text{cm}}$, and $U(r) = 2\mu V(r)/\hbar^2$ is the reduced potential. The solution must satisfy the boundary conditions
\begin{align}
u_l(r) &\to 0 \quad \text{as } r \to 0, \\
u_l(r) &\to H_l^{(-)}(\eta, kr) - S_l H_l^{(+)}(\eta, kr) \quad \text{as } r \to \infty,
\end{align}
where $H_l^{(\pm)}$ are the Coulomb-Hankel functions with Sommerfeld parameter $\eta = Z_1 Z_2 e^2 \mu / (\hbar^2 k)$, and $S_l$ is the scattering matrix element for partial wave $l$. For neutrons, $\eta = 0$ and the Coulomb functions reduce to Riccati-Hankel functions.

The nucleon-nucleus optical model potential $V(r)$ consists of real and imaginary parts describing elastic scattering and absorption, respectively. I employ the central terms of the global Koning-Delaroche (KD02) parameterization~\cite{Koning2003}, which takes the form
\begin{equation}
V(r) = V_V(r) + V_S(r) + V_C(r),
\end{equation}
where $V_V$ is the volume term, $V_S$ is the surface term, and $V_C$ is the Coulomb potential for protons. Each term has both real and imaginary components with Woods-Saxon or derivative Woods-Saxon radial forms. The spin-orbit term $V_{SO}$ is omitted as a deliberate simplification for this proof-of-concept study, since including it would require treating the $j = l \pm 1/2$ channels separately; it mainly governs spin observables such as the analyzing power and refines the angular distribution at backward angles, and its inclusion is a straightforward extension (Sec.~\ref{sec:conclusions}). The KD02 potential provides energy-dependent parameters fitted to extensive elastic scattering data for $A = 24$--$209$ nuclei at energies from 1~keV to 200~MeV. The potential depths and geometry parameters vary smoothly with energy and target mass, enabling interpolation to nuclei and energies not included in the original fit. In this work, I extend the KD02 parameterization to lighter nuclei such as \nuc{12}{C} by extrapolating the smooth mass dependence, which allows training the network over a broader range of nuclear sizes. I note that this extrapolation takes KD02 beyond its fitted range ($A \geq 24$); the results for \nuc{12}{C} and \nuc{16}{O} (both below the fitted lower bound $A = 24$) should be interpreted as a test of the network's ability to handle a wider mass range rather than as physically accurate scattering predictions for these light nuclei.

I solve Eq.~\eqref{eq:schrodinger} numerically using the Numerov method with step size $h = 0.05$~fm. The S-matrix is extracted by matching the numerical solution to Coulomb functions in the asymptotic region near $r = 15$~fm; all reported comparisons use the multi-radius average detailed in Sec.~\ref{sec:results}. Convergence has been verified by comparing with $h = 0.025$~fm and $r_{\text{match}} = 20$~fm, yielding $S$-matrix changes below $10^{-3}$ for all nuclei in the training set.

\section{Neural Network Architecture}
\label{sec:network}
\begin{figure*}[t]
\centering
\includegraphics[width=\textwidth]{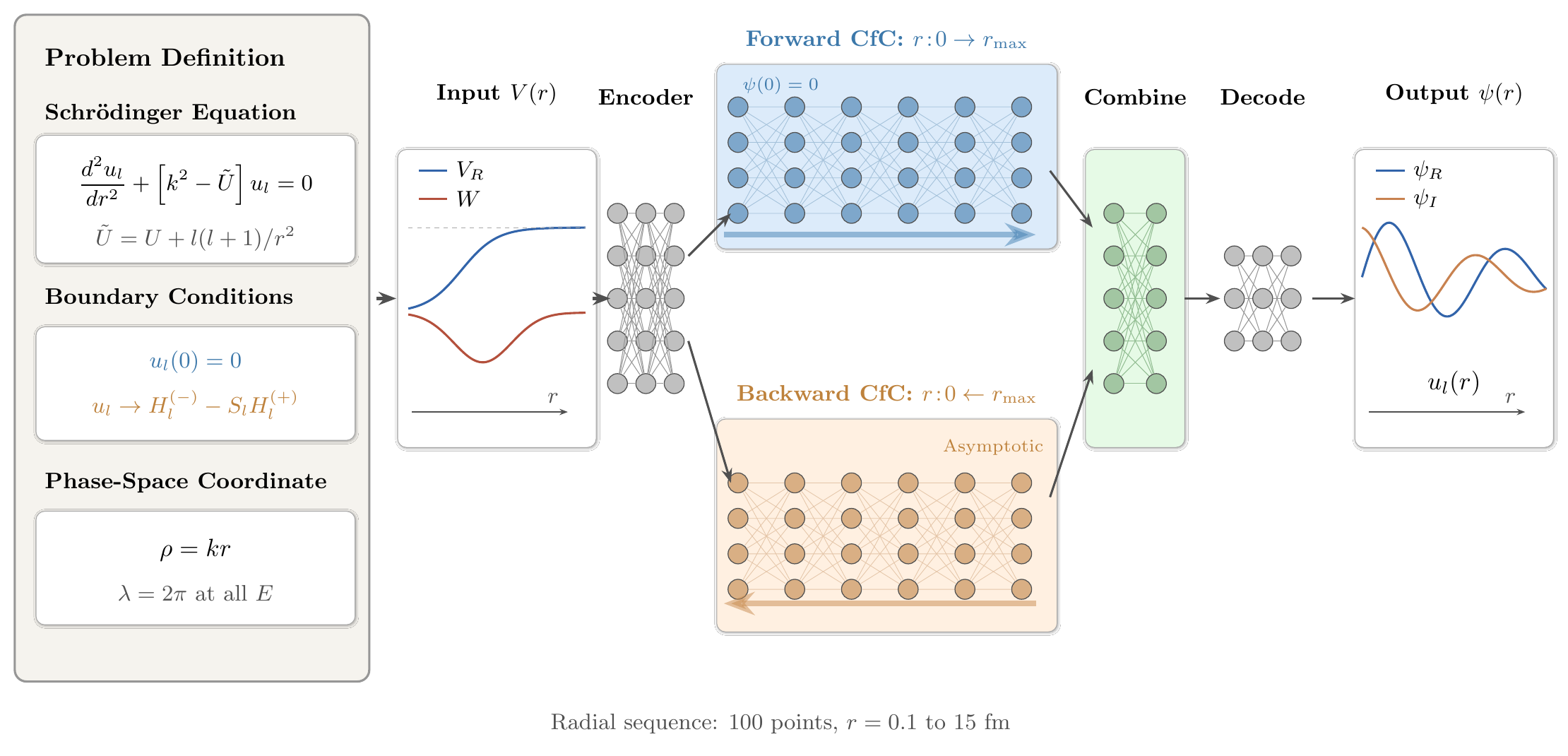}
\caption{Schematic of the Bidirectional Liquid Neural Network (BiLNN) architecture for predicting scattering wave functions. \textbf{Left panel}: The problem definition, showing the Schr\"odinger equation with effective potential $\tilde{U} = U + l(l+1)/r^2$, the boundary conditions at the origin and at infinity, and the phase-space coordinate transformation $\rho = kr$. \textbf{Main panel}: The network architecture. The input potential $V(r)$ and associated features are first processed by an encoder network. The encoded representations feed into two parallel liquid layers: the forward layer integrates from $r=0$ toward $r_{\max}$, incorporating the boundary condition $\psi(0)=0$, while the backward layer integrates from $r_{\max}$ toward $r=0$, incorporating asymptotic information. The outputs are concatenated in the combiner, then decoded to produce the real ($\psi_R$) and imaginary ($\psi_I$) parts of the radial wave function $u_l(r)$.}
\label{fig:architecture}
\end{figure*}
A neural network can be viewed as a highly flexible parameterization of a nonlinear function, conceptually similar to how a Fourier series or a set of basis functions represents a function, but with the basis functions themselves being learned from data. The basic operation is a linear transformation followed by a nonlinear function:
\begin{equation}
\mathbf{y} = \sigma(\mathbf{W}\mathbf{x} + \mathbf{b}),
\end{equation}
where $\mathbf{W}$ is a matrix of adjustable parameters (analogous to expansion coefficients), $\mathbf{b}$ is a bias vector, and $\sigma$ is a fixed nonlinear activation function, not to be confused with the scattering cross sections ($\sigma_R$, $d\sigma/d\Omega$) discussed in Sec.~\ref{sec:results}. By composing many such transformations in sequence, forming a ``deep'' network, one can represent essentially any continuous mapping~\cite{Hornik1989}. The key advantage over traditional basis expansions is that the network automatically discovers which features of the input are relevant for predicting the output.

The parameters $\mathbf{W}$ and $\mathbf{b}$ are determined by minimizing a cost function that measures the discrepancy between network predictions and known solutions. This optimization is performed using gradient descent, where the gradients are computed efficiently via the chain rule. Starting from random initial values, the parameters are iteratively updated until the cost function converges to a minimum.

For the scattering problem, the goal is to learn a mapping from the optical potential $V(r)$ and physical parameters (energy $E$, partial wave $l$, target mass $A$) to the radial wave function $u_l(r)$. In essence, the trained neural network serves as a surrogate Schr\"odinger equation solver: given the potential and boundary conditions as input, it directly outputs the wave function solution, bypassing the step-by-step numerical integration entirely. This is analogous to having a lookup table that interpolates between pre-computed solutions, but with the interpolation scheme learned automatically from the structure of the problem. Crucially, this approach learns the scattering \emph{wave function} $u_l(r)$ rather than directly predicting observables such as cross sections or $S$-matrix elements. This distinction is significant: the wave function is the fundamental quantum mechanical object from which all observables can be derived, and it encodes richer physical information than any single observable. Observable quantities like cross sections, phase shifts, and reaction rates can then be computed from the predicted wave functions using standard relations, preserving the internal consistency of quantum mechanics. This is fundamentally different from directly fitting cross section data, which would yield a purely phenomenological interpolator without access to the underlying wave function.

As noted in the Introduction, the factor of 4.5 variation in de Broglie wavelength across the 10--200~MeV range presents a challenge for a single network: wave functions at different energies have fundamentally different oscillation periods in coordinate space.

I resolve this challenge by introducing the dimensionless \emph{phase-space coordinate}
\begin{equation}
\rho = kr,
\label{eq:rho}
\end{equation}
where $k$ is the asymptotic wave number introduced in Sec.~\ref{sec:theory}. In $\rho$-space, the wavelength is always $2\pi$ regardless of the projectile energy. This transformation absorbs the energy dependence into the coordinate itself, enabling the network to learn a universal oscillation pattern that applies across all energies. The physical intuition is that $\rho$ measures distance in units of the de Broglie wavelength, so the wave function always ``looks the same'' when plotted against $\rho$. The normalization constant $\rho_{\max} = k_{\max} r_{\max} = 45.95$ is computed from the maximum energy and radial extent across the entire training set, providing a common dimensionless scale for all nuclei and energies.

For the present calculations, I discretize the radial coordinate into a sequence of 100 points uniformly spaced from $r = 0.1$~fm to $r = 15$~fm. At each point, the network receives local information about the potential and physical parameters, and must predict the corresponding wave function value. The sequence of radial points is processed as an ordered sequence, allowing the network to learn how information propagates along the radial direction, analogous to how the Numerov algorithm propagates the solution step by step.

The choice of input features significantly affects the network's ability to learn and generalize. Raw inputs such as the potential $V(r)$ and energy $E$ span vastly different numerical ranges, which can cause training difficulties. More importantly, the features should encode physically relevant information that helps the network understand the structure of the problem.

At each radial point, the network receives a 9-dimensional feature vector:
\begin{align}
\mathbf{x} = \bigg( &\frac{\rho}{\rho_{\max}}, \frac{V_R}{E}, \frac{W}{E}, \frac{\eta}{\eta_{\max}}, \sin\phi_{\text{WKB}}, \nonumber \\
&\cos\phi_{\text{WKB}}, \frac{D}{D_{\max}}, \frac{l}{l_{\max}}, \frac{A^{1/3}}{6} \bigg).
\label{eq:features}
\end{align}
Each feature is designed to capture specific physical information:

\begin{itemize}[leftmargin=*]
\item $\rho/\rho_{\max}$: The normalized phase-space coordinate, ranging from 0 to 1, tells the network ``where'' along the radial sequence it is currently located.

\item $V_R/E$ and $W/E$: The real and imaginary parts of the optical potential, normalized by the projectile energy. Here $W(r) = W_V(r) + W_S(r)$ is the total imaginary potential at each grid point, combining volume and surface contributions. The network receives the evaluated potential rather than the individual parameters, so it is agnostic to the functional form of the absorption. This ratio determines whether the particle is in a classically allowed ($V_R < E$) or forbidden ($V_R > E$) region, and how strongly absorption affects the wave function at this point.

\item $\eta/\eta_{\max}$: The Sommerfeld parameter $\eta$ (defined in Sec.~\ref{sec:theory}) characterizes the strength of the Coulomb interaction relative to the kinetic energy, normalized by a fixed reference value $\eta_{\max} = 20$ chosen to bracket the largest values encountered in the training set. For neutrons, $\eta = 0$; for protons on heavy targets at low energies, $\eta$ can be substantial.

\item $\sin\phi_{\text{WKB}}$ and $\cos\phi_{\text{WKB}}$: The WKB phase accumulated from the origin,
\begin{equation}
\phi_{\text{WKB}}(r) = \int_0^r \Re\left[ k_{\text{local}}(r') \right] dr',
\end{equation}
where $k_{\text{local}} = \sqrt{k^2 - 2\mu V(r)/\hbar^2 - l(l+1)/r^2}$ is the local wave number including the centrifugal barrier; the principal branch of the complex square root is taken with $\Im[k_{\text{local}}] \geq 0$, so that $\Re[k_{\text{local}}]$ drives the oscillatory phase $\phi_{\text{WKB}}$ and $\Im[k_{\text{local}}]$ the cumulative absorption and below-barrier decay $D(r)$. Providing both sine and cosine avoids the discontinuity at $\phi = 2\pi$ and gives the network direct information about the expected oscillation phase.

\item $D/D_{\max}$: The cumulative WKB absorption integral $D(r) = \int_0^r \Im[k_{\text{local}}(r')] \, dr'$, normalized by $D_{\max} = 734.12$ (the maximum value across the training set). This feature captures the cumulative absorption due to the imaginary potential, helping the network predict the amplitude envelope of the wave function.

\item $l/l_{\max}$, with $l_{\max} = 30$: The normalized partial wave quantum number.

\item $A^{1/3}/6$: The cube-root of the target mass number, divided by 6 so that the heaviest training target ($^{208}$Pb, $A^{1/3} \approx 5.93$) maps to approximately 1. This encodes the nuclear radius scale $R \propto A^{1/3}$ directly.
\end{itemize}

All features are normalized to lie approximately in the range $[0, 1]$ or $[-1, 1]$, which improves numerical stability during optimization. The WKB-based features ($\phi_{\text{WKB}}$ and $D$) provide semiclassical guidance, but their contribution turns out to be modest, as quantified by the ablation study in Sec.~\ref{sec:results}. This is expected because the CfC recurrent architecture inherently integrates information along the radial coordinate and can implicitly learn the WKB phase accumulation from the potential features $V/E$. The dominant factor is instead the bidirectional architecture, which incorporates asymptotic boundary information.

A basic neural network treats each radial point independently, with no mechanism to propagate information along the radial coordinate. However, the Schr\"odinger equation is a differential equation where the wave function at each point depends on its values at neighboring points, since the solution at $r$ is determined by ``integrating'' from the boundary. To capture this sequential structure, I use a \emph{recurrent} architecture that processes the radial grid points in order, maintaining an internal state $\mathbf{h}$ that accumulates information as it steps through the sequence:
\begin{equation}
\mathbf{h}_n = f(\mathbf{h}_{n-1}, \mathbf{x}_n),
\end{equation}
where $n$ indexes the radial grid points, $\mathbf{x}_n$ is the 9-dimensional feature vector [Eq.~\eqref{eq:features}] evaluated at $r_n$, and $f$ is a learned update function. This is directly analogous to how the Numerov algorithm propagates the solution from $r = 0$ outward, except that the propagation rule is learned from data rather than derived from a finite-difference approximation.

The choice of recurrent architecture requires careful consideration. Standard recurrent networks such as Long Short-Term Memory (LSTM)~\cite{Hochreiter1997} suffer from gradient degradation when processing long sequences, a significant limitation given that the radial grids contain 100 points and may need to be extended for heavy nuclei or higher energies. More fundamentally, the discrete gating mechanisms in LSTM are not naturally suited to problems governed by continuous differential equations.

I therefore employ \emph{Liquid Neural Networks}~\cite{Lechner2020,Hasani2022}, originally developed for autonomous driving and robotics, whose internal dynamics are themselves described by differential equations. This choice provides a crucial \emph{inductive bias}: rather than learning arbitrary sequence-to-sequence mappings, the network is constrained to evolve its hidden state according to a learnable ordinary differential equation, structurally aligned with the Schr\"odinger equation I seek to emulate. Specifically, the Closed-form Continuous-time (CfC) layers model the hidden state evolution as a first-order ODE with learnable time constants and nonlinear driving terms. Crucially, the ODE admits closed-form analytical solutions, providing both computational efficiency and stable gradient propagation during training.

A key innovation of the present work is the reinterpretation of liquid networks from \emph{temporal} to \emph{spatial} dynamics. The time variable is replaced with the radial coordinate $r$, transforming the network into a \emph{spatial} propagator:
\begin{equation}
\frac{d\mathbf{h}}{dr} = -\frac{\mathbf{h}}{\Lambda} + \mathbf{g}(\mathbf{h}, \mathbf{x}),
\end{equation}
where $\Lambda$ now represents a learnable ``relaxation length'' rather than a time constant. This reinterpretation is natural for the scattering problem: just as the original CfC layers learn to propagate information forward in time, the spatial liquid layers learn to propagate the wave function solution along the radial coordinate, in a manner directly analogous to numerical integration of the Schr\"odinger equation.

The scattering Schr\"odinger equation is a \emph{boundary value problem} with conditions specified at both ends of the domain: $u_l(0) = 0$ at the origin and the asymptotic Coulomb behavior $u_l \to H_l^{(-)} - S_l H_l^{(+)}$ as $r \to \infty$. A unidirectional network that processes only from small to large $r$ naturally incorporates information about the left boundary condition but has no direct knowledge of the right boundary condition, since it can only ``see'' what came before, not what comes after.

To incorporate both boundary conditions, I employ a \emph{bidirectional} architecture with two parallel networks. The \textbf{forward} network processes the radial sequence from the innermost grid point ($r = 0.1$~fm, where $u_l$ is already negligibly small) outward to $r_{\max}$, propagating information about the origin boundary condition $u_l(0) = 0$. The \textbf{backward} network processes the sequence in reverse, from $r_{\max}$ back to the innermost point, propagating information about the asymptotic behavior inward. At each radial point, the outputs from both directions are combined to form a representation that incorporates information from both boundaries. This bidirectional processing is analogous to the shooting method in numerical analysis, where solutions are propagated from both boundaries and matched in the middle.

The complete BiLNN architecture consists of five components (see Fig.~\ref{fig:architecture}): (i) an \textbf{encoder} that transforms the 9-dimensional input features into a higher-dimensional internal representation suitable for processing; (ii) the \textbf{forward liquid} layer that processes this representation from the innermost grid point outward; (iii) the \textbf{backward liquid} layer that processes in reverse from $r_{\max}$ inward; (iv) a \textbf{combiner} that merges the forward and backward information at each radial point; and (v) a \textbf{decoder} that transforms the internal representation back into the physical output $(\psi_R, \psi_I)$. The liquid layers use a structured sparse connectivity (AutoNCP wiring~\cite{Lechner2020}) with 128 internal units each.

The detailed layer specification is as follows. The \textbf{encoder} consists of three fully connected layers with SiLU activation: $9 \to 512 \to 512 \to 512$. Each CfC liquid layer has 128 internal units with AutoNCP wiring and produces a 64-dimensional output at each radial point. The \textbf{combiner} concatenates the forward and backward outputs ($64 + 64 = 128$ dimensions) and processes them through two layers: $128 \to 128 \to 64$. The \textbf{decoder} maps this to the wave function via four layers: $64 \to 64 \to 32 \to 16 \to 2$, where the final two outputs are $(\psi_R, \psi_I)$.

The total model has approximately 1,290,922 adjustable parameters. While this may seem large, these parameters are determined once during training and then fixed. The trained network can evaluate wave functions for any combination of inputs within its training domain essentially instantaneously, since the computational cost of solving the differential equation has been amortized into the one-time training process.

Training data is generated by solving Eq.~\eqref{eq:schrodinger} numerically using the Numerov method for random combinations of physical parameters. The training set covers:
\begin{itemize}[leftmargin=*]
\item Twelve target nuclei: \nuc{12}{C}, \nuc{16}{O}, \nuc{27}{Al}, \nuc{28}{Si}, \nuc{40}{Ca}, \nuc{48}{Ti}, \nuc{56}{Fe}, \nuc{58}{Ni}, \nuc{90}{Zr}, \nuc{120}{Sn}, \nuc{197}{Au}, and \nuc{208}{Pb}, selected to provide representative coverage across the range of nuclear radii ($\propto A^{1/3}$), spanning light, medium-mass, and heavy systems.
\item Both projectile types: protons and neutrons.
\item Energies randomly sampled from a uniform distribution over 1 to 200~MeV.
\item Partial waves from $l = 0$ to $l = 30$, sufficient to converge the cross section at all energies considered.
\end{itemize}

For each combination of target and projectile type, I generate 200 random energy samples, drawing each energy independently from $U(1, 200)$~MeV. Critically, all 31 partial waves ($l = 0, \ldots, 30$) are computed at each sampled energy, ensuring that a complete set of partial waves at a single energy is available for cross-section calculation. This yields approximately 148,800 training examples in total. The Numerov solutions, computed on a fine grid with step $h = 0.05$~fm, are interpolated (using linear interpolation) onto the 100-point uniform grid for network training. The wave functions are then normalized by their maximum absolute value, $\max|u_l(r)| = 1$, to bring all samples to a comparable scale; the $S$-matrix extraction via asymptotic matching (Sec.~\ref{sec:results}) is normalization-independent, so this choice does not affect observables. Each wave function is evaluated on the uniform grid of 100 radial points, spaced from $r = 0.1$ to 15~fm with step $\Delta r \approx 0.15$~fm. The normalization constants $\rho_{\max}$ and $D_{\max}$ introduced above are computed once from the full training set (random seed 42) and held fixed during inference.

The dataset is split into 80\% for training and 20\% for testing, with the split performed over individual (target, energy, partial-wave) samples. The test set is used to evaluate generalization performance on parameter combinations not seen during training. Importantly, the test set includes interpolation within the training domain (e.g., an energy of 75~MeV when trained on samples at 70 and 80~MeV) rather than extrapolation beyond it. Because all partial waves at a given sampled energy are generated together, this held-out set probes interpolation within the training domain rather than fully independent energy points; the genuine out-of-distribution test is provided by the three target nuclei (\nuc{24}{Mg}, \nuc{63}{Cu}, \nuc{184}{W}) excluded from training entirely [Fig.~\ref{fig:generalization}].

The cost function to be minimized is the mean squared error between predicted and true wave functions, where $\psi_R$ and $\psi_I$ denote the real and imaginary parts of the complex wave function $u_l(r)$:
\begin{equation}
\mathcal{L} = \frac{1}{N} \sum_{i=1}^{N} \left[ (\psi_{R,i}^{\text{pred}} - \psi_{R,i}^{\text{true}})^2 + (\psi_{I,i}^{\text{pred}} - \psi_{I,i}^{\text{true}})^2 \right],
\end{equation}
where the sum runs over all radial points in all training samples. This treats the real and imaginary parts equally and penalizes deviations at all radii uniformly.

The optimization is performed using the AdamW algorithm~\cite{Loshchilov2019} with initial learning rate $10^{-3}$, weight decay $10^{-4}$, and batch size 512. The learning rate is reduced over 500 epochs using a cosine annealing schedule. Gradient norms are clipped to 1.0 to stabilize training. Mixed-precision (float16) training is employed on CUDA-capable GPUs via PyTorch's automatic mixed precision (AMP), reducing memory usage and accelerating computation without measurable loss of accuracy. On a modern GPU (NVIDIA RTX 5090), training completes in several hours. The implementation uses PyTorch~2.0+ and the \texttt{ncps} package~\cite{Lechner2020,Hasani2022} for CfC layers. Code is available from the author upon reasonable request.

\begin{figure}[t]
\centering
\includegraphics[width=\columnwidth]{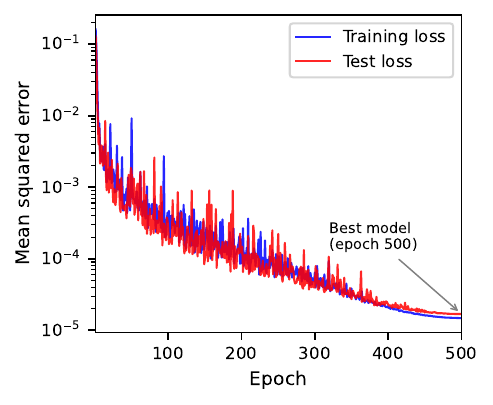}
\caption{Training convergence of the baseline BiLNN model. Both training (blue) and test (red) mean squared error decrease by four orders of magnitude over 500 epochs. The close overlap of the two curves indicates the absence of overfitting.}
\label{fig:convergence}
\end{figure}

Figure~\ref{fig:convergence} shows the training and test loss, defined as the mean-squared error of the predicted wave function on the held-out test set, as a function of epoch. The loss decreases rapidly during the first $\sim$100 epochs (from $\sim 0.1$ to $\sim 10^{-3}$), then enters a slow-refinement regime; by $\sim$epoch 350 the test loss has largely flattened, with each additional epoch reducing it only marginally. The training and test curves remain closely overlapping throughout, indicating that the model is not overfitting. The training length of 500 epochs was fixed on this basis: it lies within the plateau of the monitored test loss, where further training reduces the loss negligibly and leaves the predicted observables essentially unchanged. I verified this directly by training a separate model with a longer cosine-annealing schedule (700 epochs); this lowers the test error by less than $0.1$ percentage point and leaves the $S$-matrix and cross sections unchanged.

\section{Results}
\label{sec:results}
\begin{figure}[t]
\centering
\includegraphics[width=\columnwidth]{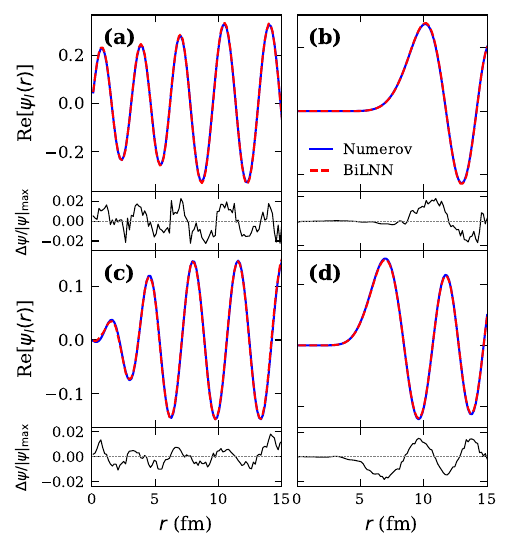}
\caption{Comparison of wave functions predicted by BiLNN (dashed red lines) with numerically exact Numerov solutions (solid blue lines) for representative test cases. (a) p+\nuc{208}{Pb} at $E=75$~MeV, $l=0$: $s$-wave showing full penetration to the nuclear interior with $\sim$5 oscillations; (b) p+\nuc{208}{Pb} at $E=75$~MeV, $l=15$: high-$l$ partial wave with centrifugal barrier suppression at small $r$; (c) n+\nuc{40}{Ca} at $E=69$~MeV, $l=0$: neutron $s$-wave without Coulomb effects; (d) n+\nuc{40}{Ca} at $E=69$~MeV, $l=10$: centrifugal suppression at intermediate energy. Lower sub-panels show the point-wise residual $\Delta\psi/\max|\psi^{\mathrm{true}}|$, confirming sub-percent agreement across the entire radial range.}
\label{fig:wavefunction}
\end{figure}
The overall wave function error is defined as the mean absolute deviation between the predicted and reference wave functions, normalized by the mean reference amplitude,
\begin{equation}
\epsilon = \frac{\sum_{i,n} |\psi_i^{\text{pred}}(r_n) - \psi_i^{\text{true}}(r_n)|}{\sum_{i,n} |\psi_i^{\text{true}}(r_n)|} \times 100\%,
\label{eq:error}
\end{equation}
where the sums run over all test samples $i$ and all radial grid points $r_n$.

\begin{figure}[t]
\centering
\includegraphics[width=\columnwidth]{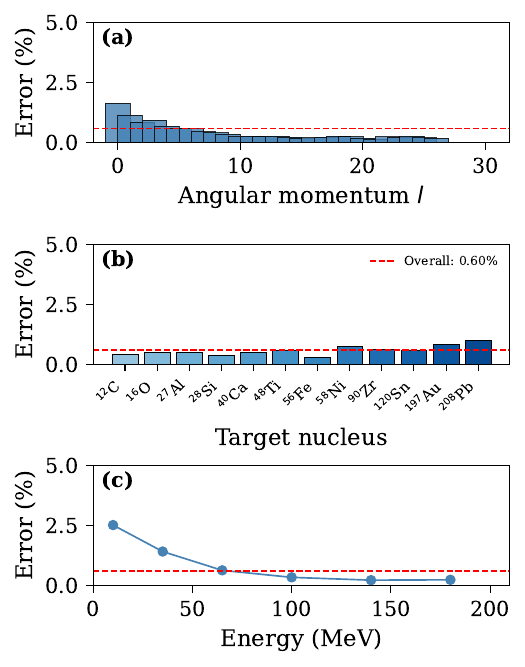}
\caption{Summary of $S$-matrix prediction errors for physically contributing partial waves ($l < kR + 5$, where $R \approx 1.2 A^{1/3}$~fm). (a) Error versus partial wave $l$, showing elevated errors ($\sim$1.0--1.3\%) at low $l$ (deep nuclear penetration), decreasing to $\lesssim$0.5\% at higher $l$; (b) Error by target nucleus for all twelve training targets, with uniform accuracy (0.3--0.8\%) across most nuclei and slightly higher errors ($\sim$1\%) for the heaviest targets; (c) Error versus projectile energy, showing the strongest systematic dependence: $\sim$2.5\% at low energies (10--20~MeV) decreasing to $\lesssim$0.5\% above 50~MeV. The dashed lines indicate the overall average $S$-matrix error of 0.6\%.}
\label{fig:errors}
\end{figure}

The trained BiLNN is evaluated on a held-out test set comprising 20\% of the total data (approximately 29,760 samples), with parameter combinations not seen during training. The overall relative error, defined by Eq.~\eqref{eq:error} and averaged over all test samples, is 0.6\%. This pooled figure is energy dependent: the error falls below 0.5\% above 50~MeV and rises to about 2.5\% in the 10--20~MeV band (Fig.~\ref{fig:errors}), reflecting the longer wavelengths and stronger relative Coulomb effects at low energy. The corresponding $S$-matrix error, computed using the matching method averaged over 15 matching radii in the asymptotic region (a multi-point average that suppresses derivative noise, detailed below), is also approximately 0.6\%: the matching extraction is a ratio that is independent of the overall normalization, so it preserves rather than amplifies the wave function accuracy.

Figure~\ref{fig:wavefunction} presents a detailed comparison of wave functions predicted by BiLNN with numerically exact Numerov solutions for four representative test cases that span different projectile types, target nuclei, energies, and partial waves. Panel~(a) shows the $s$-wave ($l=0$) radial wave function for proton scattering on \nuc{208}{Pb} at $E = 75$~MeV, where the real part exhibits approximately 5 complete oscillations within the radial range $r = 0$--15~fm, with amplitude growing from small values near the origin (where the strong nuclear and Coulomb potentials suppress the wave function) to maximum amplitude around $r \approx 8$~fm before settling into the asymptotic Coulomb behavior. The BiLNN prediction (dashed red) overlays the Numerov reference (solid blue) with a point-wise relative error of 0.8\% for this case. Panel~(b) displays the $l=15$ partial wave for the same p+\nuc{208}{Pb} system at $E = 75$~MeV. Here the centrifugal barrier $l(l+1)/r^2$ creates a classically forbidden region at small radii, suppressing the wave function to nearly zero for $r \lesssim 5$~fm. Beyond the classical turning point, the wave function emerges with oscillatory behavior confined to the exterior region $r > 5$~fm. The network accurately captures both the exponential suppression in the forbidden region and the subsequent oscillatory structure, including the correct phase at the matching radius. Panels~(c) and (d) show neutron scattering on \nuc{40}{Ca}, where the absence of Coulomb interaction simplifies the potential but the wave function must still satisfy the correct asymptotic matching to spherical Bessel functions. At $E = 69$~MeV [panel~(c), $l=0$], the $s$-wave penetrates to the origin and shows approximately 4 oscillations with the characteristic pattern of increasing amplitude toward the asymptotic region. At $E = 69$~MeV [panel~(d), $l=10$], the centrifugal barrier produces a wave function that is suppressed for $r \lesssim 4$~fm and exhibits oscillatory behavior in the exterior region. In all cases, the point-wise agreement between BiLNN and Numerov is consistently below 1\% across the entire radial range, as directly visible in the residual sub-panels of Fig.~\ref{fig:wavefunction}, demonstrating that the phase-space coordinate representation successfully enables generalization across the wide energy range.

\begin{figure}[t]
\centering
\includegraphics[width=\columnwidth]{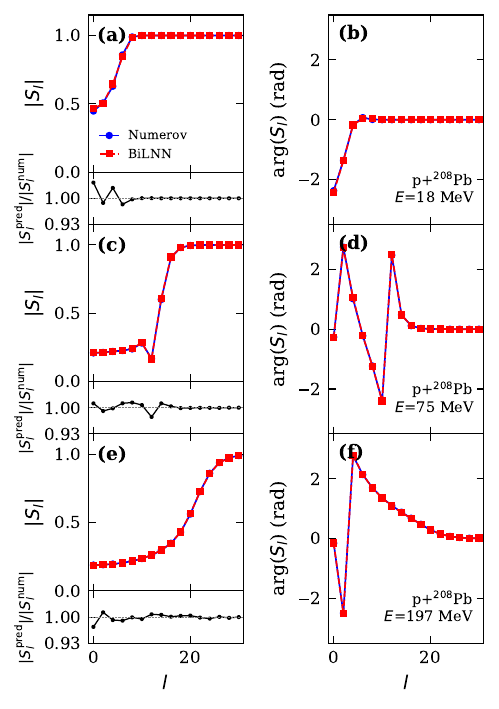}
\caption{$S$-matrix comparison for p+\nuc{208}{Pb} scattering at three energies, arranged in rows from top to bottom: $E \approx 18$~MeV, $E \approx 75$~MeV, and $E \approx 197$~MeV. Left panels [(a), (c), (e)]: $|S_l|$ showing oscillatory absorption at low $l$ with sharp transition to transparency ($|S_l| \to 1$) at the grazing angular momentum $l_g$, which increases with energy from $l_g \approx 8$ at 18~MeV to $l_g \approx 22$ at 197~MeV. Below each $|S_l|$ panel, a ratio sub-panel displays $|S_l^{\mathrm{pred}}|/|S_l^{\mathrm{Numerov}}|$, confirming agreement within $\pm 2\%$ across all partial waves. Right panels [(b), (d), (f)]: $\arg(S_l)$ showing the nuclear and Coulomb phase shifts. Blue circles: Numerov reference; red squares: BiLNN prediction.}
\label{fig:smatrix}
\end{figure}

Since the ultimate goal of optical model calculations is to predict scattering observables, I verify that the wave function accuracy translates to accurate $S$-matrix elements and cross sections. The $S$-matrix is extracted from the predicted wave functions using the asymptotic matching method~\cite{Thompson2009}
\begin{equation}
S_l = \frac{H_l^{(-)'} \psi_l - H_l^{(-)} \psi_l'}{H_l^{(+)'} \psi_l - H_l^{(+)} \psi_l'},
\label{eq:smatrix_matching}
\end{equation}
where $H_l^{(\pm)}$ are the Coulomb-Hankel functions and $\psi_l$, $\psi_l'$ are the wave function and its derivative evaluated at the matching radius. This ratio formulation is preferred because it is independent of the wave function normalization: any arbitrary scaling factor cancels between numerator and denominator. This property is essential because the neural network is trained on wave functions normalized to $\max|\psi_l(r)| = 1$ for numerical stability, rather than the physical normalization $\psi_l \to (H_l^{(-)} - S_l H_l^{(+)})/(2ik)$ as $r \to \infty$. The matching method extracts the correct $S$-matrix regardless of this arbitrary normalization choice.

To reduce numerical noise from the derivative estimation at a single radius, I average the extracted $S$-matrix over 15 consecutive matching radii in the asymptotic region ($r = 12$--15~fm), where the nuclear potential has decayed to negligible values. This multi-point averaging significantly improves numerical stability without introducing systematic bias, and the identical extraction, including the numerical derivative of the wave function, is applied to both the BiLNN and Numerov solutions, since all radii in this region should yield identical $S$-matrix values for the exact solution.

The error propagation from wave function to observables exhibits an important hierarchy. The 0.6\% wave function error is preserved at the $S$-matrix level ($\sim$0.6\%), since the matching method involves ratios that cancel the overall normalization. However, the differential cross section $d\sigma/d\Omega$ involves a coherent sum over $\sim$30 partial waves,
\begin{equation}
\frac{d\sigma}{d\Omega} = \left| f_C(\theta) + \sum_l \frac{2l+1}{2ik} (S_l - 1) P_l(\cos\theta) e^{2i\sigma_l} \right|^2,
\end{equation}
where $f_C(\theta)$ is the Coulomb (Rutherford) amplitude, $P_l$ the Legendre polynomial, $\sigma_l$ the Coulomb phase shift, and small phase errors in individual $S_l$ can accumulate. At forward angles where constructive interference dominates, cross-section errors remain modest ($\sim$1--10\%). At backward angles ($\theta \approx 180°$), where the Coulomb amplitude $f_C$ and nuclear amplitude $f_N$ have comparable magnitudes but opposite phases, destructive interference leads to catastrophic cancellation: the total amplitude $|f_C + f_N| \ll |f_C|, |f_N|$ becomes highly sensitive to small phase errors, amplifying the cross-section error at the deepest minima to as much as $\sim$50\%. This is a fundamental numerical sensitivity of backward-angle scattering, not a limitation specific to the neural network. Integral quantities such as the reaction cross section $\sigma_R = \pi\lambda^2 \sum_l (2l+1)(1-|S_l|^2)$ (with $\lambda = 1/k$ the reduced wavelength), which do not involve coherent interference, remain accurate to well below 1\%.

\begin{figure*}[t]
\centering
\includegraphics[width=\textwidth]{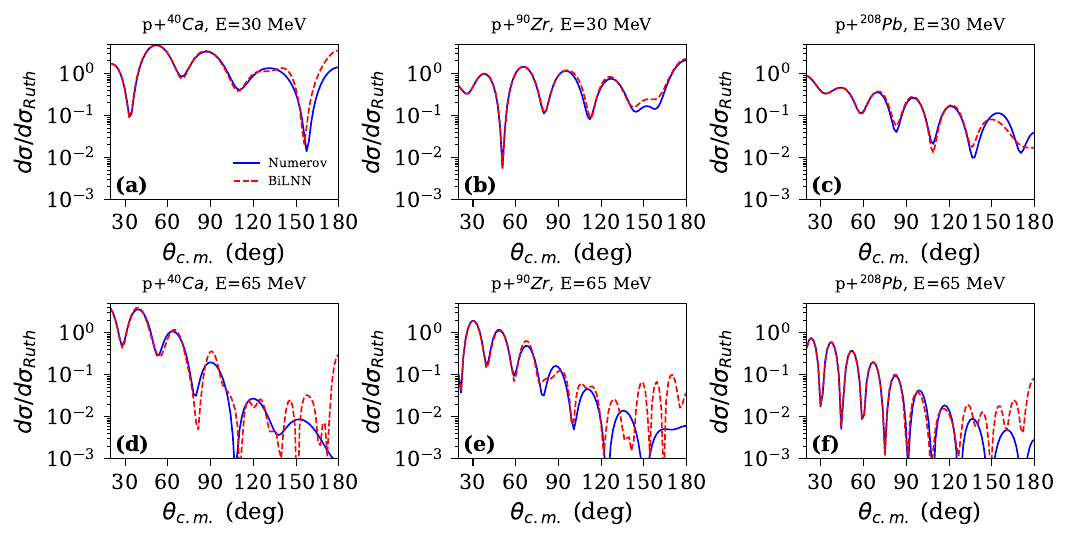}
\caption{Proton elastic scattering angular distributions $d\sigma/d\sigma_{\text{Ruth}}$ for three target nuclei at two energies each. (a,d) \nuc{40}{Ca} at $E = 30$~MeV and $E = 65$~MeV; (b,e) \nuc{90}{Zr} at $E = 30$~MeV and $E = 65$~MeV; (c,f) \nuc{208}{Pb} at $E = 30$~MeV and $E = 65$~MeV. The top row shows lower-energy cases with broader diffraction features; the bottom row shows higher-energy cases with more compressed, sharper minima. Blue solid lines: Numerov reference; red dashed lines: BiLNN prediction. The diffraction patterns, including positions, depths, and widths of multiple minima, are reproduced across up to four orders of magnitude in cross section ratio.}
\label{fig:proton_xsec}
\end{figure*}

\begin{figure*}[t]
\centering
\includegraphics[width=\textwidth]{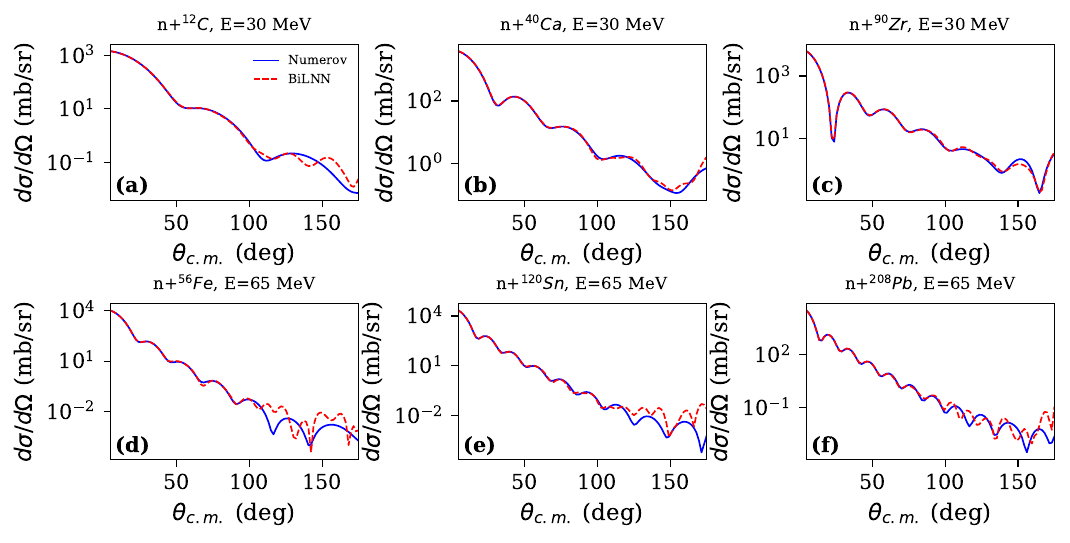}
\caption{Neutron elastic scattering angular distributions $d\sigma/d\Omega$ for six target nuclei spanning the training set. (a) \nuc{12}{C} at $E = 30$~MeV; (b) \nuc{40}{Ca} at $E = 30$~MeV; (c) \nuc{90}{Zr} at $E = 30$~MeV; (d) \nuc{56}{Fe} at $E = 65$~MeV; (e) \nuc{120}{Sn} at $E = 65$~MeV; (f) \nuc{208}{Pb} at $E = 65$~MeV. Blue solid lines: Numerov reference; red dashed lines: BiLNN prediction.}
\label{fig:neutron_xsec}
\end{figure*}

Figure~\ref{fig:smatrix} compares the $S$-matrix modulus $|S_l|$ and phase $\arg(S_l)$ obtained from BiLNN and Numerov wave functions for proton scattering on \nuc{208}{Pb} at three representative energies: 18~MeV, 75~MeV, and 197~MeV, arranged in rows from top to bottom. For each energy, the left panels display $|S_l|$ as a function of partial wave $l$ with a ratio sub-panel $|S_l^{\mathrm{pred}}|/|S_l^{\mathrm{Numerov}}|$ underneath, and the right panels show $\arg(S_l)$. At the lowest energy $E = 18$~MeV [panel~(a)], the $S$-matrix modulus shows a complex oscillatory pattern at low $l$, with $|S_l|$ varying between 0.4 and 0.9 for $l \lesssim 8$, reflecting interference between nuclear and Coulomb amplitudes in this strongly absorbing regime. Beyond $l \approx 8$, the transition to transparency ($|S_l| \approx 1$) is sharp, occurring over just 2--3 partial waves. At intermediate energy $E = 75$~MeV [panel~(c)], the absorption region extends to higher partial waves ($l \lesssim 14$), with $|S_l|$ showing oscillatory structure in the interior region before the sharp rise to unity. At the highest energy $E = 197$~MeV [panel~(e)], the grazing angular momentum shifts further to $l_g \approx 22$, and the oscillatory pattern in $|S_l|$ becomes more pronounced, with values dropping as low as 0.2 before recovering. This energy-dependent shift of the absorption edge reflects the increasing classical impact parameter $b = l/k$ at higher energies, with the grazing condition $b \approx R$ determining where the projectile trajectory begins to miss the nuclear volume. The BiLNN predictions (red squares) overlay the Numerov reference values (blue circles) agreeing within 0.5\% for strongly absorbed waves ($l < l_g$) and 0.3\% for peripheral waves ($l > l_g$), accurately capturing the oscillatory structure, absorption depth, and transition sharpness. The ratio sub-panels beneath each $|S_l|$ plot quantitatively confirm this agreement, with $|S_l^{\mathrm{pred}}|/|S_l^{\mathrm{Numerov}}|$ clustering tightly around unity across all partial waves and energies. The right panels [(b), (d), (f)] show the $S$-matrix phase $\arg(S_l)$, which encodes the combined nuclear and Coulomb phase shifts. At low energies [panel~(b)], the phase exhibits rapid variation with $l$, starting near $+2$ radians at $l=0$ and dropping sharply to $-2$ radians around $l \approx 6$ before recovering to zero at high $l$. At higher energies [panels (d) and (f)], the phase variation becomes smoother but extends to higher $l$ values before asymptoting to zero. The complex $S$-matrix $S_l = |S_l|\,e^{2i\delta_l}$ is reproduced with a sub-percent relative error $|S_l^{\mathrm{pred}} - S_l^{\mathrm{Numerov}}|/|S_l^{\mathrm{Numerov}}|$, so both the modulus and the phase are captured. This is crucial since even small phase errors would produce significant deviations in the interference patterns of the cross section, as confirmed by the cross-section comparisons in Figs.~\ref{fig:proton_xsec}--\ref{fig:neutron_xsec}.

The elastic scattering cross section is the experimentally observable quantity that tests whether the coherent sum over partial waves preserves the accuracy of the individual $S$-matrix elements. Figure~\ref{fig:proton_xsec} shows the proton elastic scattering angular distribution expressed as the ratio to Rutherford scattering, $d\sigma/d\sigma_{\text{Ruth}}$, for three target nuclei at two energies each: \nuc{40}{Ca} [panels (a) and (d)], \nuc{90}{Zr} [panels (b) and (e)], and \nuc{208}{Pb} [panels (c) and (f)]. The top row shows lower-energy cases, while the bottom row shows higher-energy cases. In all cases, the BiLNN predictions (red dashed lines) faithfully reproduce both the positions and depths of the diffraction minima. At backward angles ($\theta > 120°$), discrepancies of 10--20\% are visible (reaching $\sim$50\% at the deepest diffraction minima), particularly for heavy targets at lower energies. These arise from the sensitivity of backward-angle scattering to the coherent cancellation between Coulomb and nuclear amplitudes, which amplifies small phase errors in low partial waves.

For neutron scattering, where no Coulomb interaction is present, Figure~\ref{fig:neutron_xsec} displays the differential cross section $d\sigma/d\Omega$ in mb/sr for six target nuclei: \nuc{12}{C} [panel (a)], \nuc{40}{Ca} [panel (b)], \nuc{90}{Zr} [panel (c)], \nuc{56}{Fe} [panel (d)], \nuc{120}{Sn} [panel (e)], and \nuc{208}{Pb} [panel (f)]. The absence of the Coulomb amplitude eliminates the Rutherford divergence at forward angles. In all cases, the BiLNN predictions capture both the overall magnitude and the detailed oscillatory structure. This does not mean neutron scattering is free of the sensitivity discussed above: it lacks the specific Coulomb-nuclear cancellation that drives the proton backward-angle errors, but the same amplification of small phase errors persists at the deep diffraction minima, where the cross section is smallest and the coherent partial-wave sum nearly cancels.

Figure~\ref{fig:errors} provides a comprehensive summary of the $S$-matrix error distribution across the parameter space, restricted to physically contributing partial waves satisfying $l < kR + 5$ where $R \approx 1.2 A^{1/3}$~fm. This criterion excludes high-$l$ partial waves for light nuclei at low energies, where the centrifugal barrier prevents the wave function from penetrating into the nuclear interior and therefore makes negligible contribution to the cross section. The $S$-matrix error is computed as the relative error in $|S_l|$ between BiLNN and Numerov, averaged over 15 matching radii to ensure numerical stability. Panel~(a) displays the relative error as a function of partial wave quantum number $l$. The error shows a non-monotonic dependence on angular momentum: low-$l$ partial waves ($l \lesssim 4$) exhibit the largest errors ($\sim$1.0--1.3\%), while higher partial waves achieve progressively lower errors ($\lesssim$0.5\%). The elevated error at low $l$ can be understood physically: for $s$-waves and low partial waves, the wave function penetrates deep into the nuclear interior where both the nuclear attraction and Coulomb repulsion (for protons) are strongest, requiring the network to capture rapid variations in a region of steep potential gradients. Panel~(b) shows the error breakdown by target nucleus for all twelve nuclei in the training set. The errors are uniform across most nuclei, clustering around 0.3--0.8\%, with the heaviest targets \nuc{197}{Au} and \nuc{208}{Pb} showing slightly larger errors around 1\%. This trend can be attributed to the stronger Coulomb interaction for proton scattering on heavy targets, which introduces additional oscillatory structure at large distances and requires more extended matching radii. Notably, \nuc{56}{Fe} shows the lowest error among the mid-mass nuclei, consistent with its position in the iron-peak region where the KD02 parameters vary most smoothly with $A$ and $Z$. The horizontal dashed line indicates the overall average $S$-matrix error of 0.6\%. Panel~(c) reveals the most pronounced systematic dependence: the error varies significantly with projectile energy, from approximately 2.5\% at the lowest energies (10--20~MeV) down to about 0.2--0.5\% at high energies (above 50~MeV). This trend reflects several compounding difficulties at low energies. First, the ratio $V_R/E$ becomes very large inside the nucleus, pushing the potential input features to extreme values where the network has fewer training examples and must extrapolate in feature space. Second, for proton scattering, the Sommerfeld parameter $\eta \propto 1/\sqrt{E}$ grows substantially, introducing long-range Coulomb oscillations that extend well beyond the nuclear radius and require the network to maintain coherence over the full radial domain. Third, the wave function exhibits fewer oscillation cycles within the nuclear volume, providing less spatial structure for the network to distinguish and match phase. Although the $\rho = kr$ transformation successfully normalizes the oscillation wavelength (all wave functions oscillate with period $2\pi$ in $\rho$-space), it does not address these additional sources of low-energy difficulty. At higher energies, the more rapid oscillations are well captured by the phase-space coordinate representation, the $V_R/E$ ratios are moderate, and $\eta$ is small, resulting in $S$-matrix errors well below 1\%.

\begin{figure*}[t]
\centering
\includegraphics[width=\textwidth]{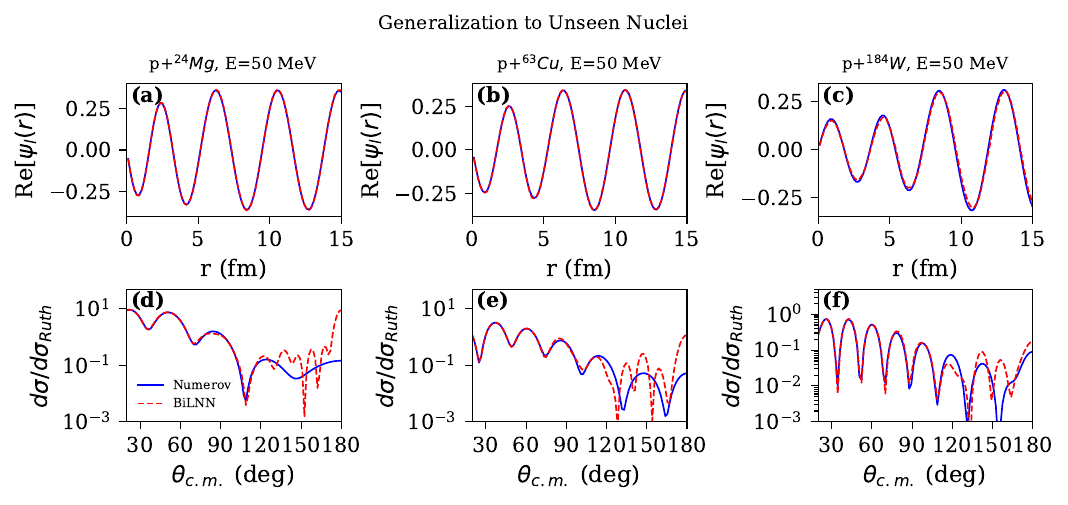}
\caption{Generalization test for nuclei \emph{not} included in training: \nuc{24}{Mg} (left), \nuc{63}{Cu} (center), and \nuc{184}{W} (right). Top row [(a)--(c)]: $s$-wave proton wave functions at $E = 50$~MeV, showing the real part of $\psi_l(r)$; bottom row [(d)--(f)]: corresponding elastic scattering cross sections $d\sigma/d\sigma_{\text{Ruth}}$ spanning up to four orders of magnitude. Blue: Numerov reference; red: BiLNN prediction. The network generalizes successfully to these unseen nuclei: the wave-function error, averaged over all energies and partial waves, is 0.51\%, 0.55\%, and 0.86\% for \nuc{24}{Mg}, \nuc{63}{Cu}, and \nuc{184}{W} respectively, comparable to the 0.6\% in-sample accuracy. This demonstrates that the network has captured the smooth functional dependence encoded in the KD02 parameterization rather than memorizing target-specific patterns.}
\label{fig:generalization}
\end{figure*}

A key test of whether the network has captured the systematic trends in the KD02-Numerov solutions rather than merely memorizing the training data is its ability to generalize to nuclei not included in training. Figure~\ref{fig:generalization} shows predictions for three nuclei that were never seen during training: \nuc{24}{Mg} ($A=24$, $Z=12$), \nuc{63}{Cu} ($A=63$, $Z=29$), and \nuc{184}{W} ($A=184$, $Z=74$). These targets were deliberately chosen to lie between the training nuclei in atomic mass: \nuc{24}{Mg} between \nuc{16}{O} and \nuc{27}{Al}, \nuc{63}{Cu} between \nuc{58}{Ni} and \nuc{90}{Zr}, and \nuc{184}{W} between \nuc{120}{Sn} and \nuc{197}{Au}, testing interpolation capability with nuclei having different proton-to-neutron ratios that affect the optical potential parameters through the isospin dependence of the KD02 parameterization. The top row [panels (a)--(c)] displays $s$-wave ($l=0$) proton scattering wave functions at $E = 50$~MeV. For \nuc{24}{Mg} [panel~(a)], the light target produces a wave function with approximately 4 oscillations and relatively uniform amplitude, characteristic of weak absorption in the small nuclear volume. For \nuc{63}{Cu} [panel~(b)], the larger nuclear radius shifts the oscillatory pattern outward with approximately 5 oscillations visible, while the wave function amplitude grows more pronounced toward the asymptotic region. For \nuc{184}{W} [panel~(c)], the heavy target produces a wave function with similar oscillation count but distinct amplitude pattern due to stronger absorption, with sub-percent agreement between BiLNN and Numerov across the full radial range. The bottom row [panels (d)--(f)] shows the corresponding elastic scattering cross sections $d\sigma/d\sigma_{\text{Ruth}}$. For \nuc{24}{Mg} [panel~(d)], the cross section ratio remains above unity across most angles with a single broad minimum. For \nuc{63}{Cu} [panel~(e)], multiple diffraction minima appear spanning nearly four orders of magnitude in cross section ratio. For \nuc{184}{W} [panel~(f)], the strong Coulomb field produces a pattern with the ratio declining smoothly from forward angles to backward angles. In all cases, the BiLNN predictions closely track the Numerov reference, demonstrating that the network has captured the smooth dependence of the Numerov solution on atomic mass $A$ and charge $Z$ as encoded in the KD02 parameterization, rather than memorizing target-specific patterns. Averaged over all energies and partial waves, the wave-function error for these unseen nuclei is 0.51\%, 0.55\%, and 0.86\% for \nuc{24}{Mg}, \nuc{63}{Cu}, and \nuc{184}{W} respectively, comparable to the 0.6\% in-sample test error; the heaviest target, \nuc{184}{W}, shows the largest error, following the same trend with target mass observed for the training nuclei.

The bidirectional architecture proves essential for achieving this level of accuracy. I compared the BiLNN with a unidirectional variant that processes the radial sequence only in the forward direction (from $r=0$ outward). On the same training data, the unidirectional network achieved an overall error of 0.81\%, compared to 0.61\% for the bidirectional version, a relative degradation of 33\% with respect to the bidirectional baseline. The impact is most severe for heavy nuclei ($^{208}$Pb: 1.09\% vs 0.75\%) and at low energies (1--20~MeV: 2.76\% vs 1.65\%), where the backward pass provides essential information about the asymptotic boundary condition that propagates inward to constrain the interior solution.

To validate the design choices systematically, I performed a comprehensive ablation study in which each architectural and data component was varied independently while keeping all other settings fixed. Table~\ref{tab:ablation} summarizes the results. The bidirectional architecture proves most critical: as the dedicated comparison above shows, removing the backward CfC layer is the single most damaging change. The WKB input features provide only a modest improvement (removing them changes the error from 0.61\% to 0.62\%), while reducing to a minimal 4-feature input degrades accuracy to 0.75\%. The 128-unit CfC model offers the best accuracy--efficiency trade-off; a larger 256-unit model achieves 0.53\% error but with 56\% more parameters. The training set size matters: reducing from 12 to 6 nuclei increases the error from 0.6\% to 0.73\%, with the largest degradation for heavy nuclei. The 100-point radial grid is optimal; interestingly, a finer 200-point grid performs slightly worse (0.72\%). This counterintuitive result reflects a fundamental trade-off in recurrent architectures: doubling the sequence length doubles the number of recurrent steps through which gradient signals must propagate during backpropagation, exacerbating the vanishing gradient problem and making it harder for the CfC layers to maintain long-range correlations across the full radial domain. In the scattering problem, the network must propagate boundary condition information from both ends of the domain ($r=0$ and $r=r_{\max}$) to the nuclear interior, a distance spanning the entire sequence. With 200 points, this requires twice as many recurrent steps as with 100 points, while the physical information content per step (the potential variation over $\Delta r \approx 0.075$~fm vs.\ $0.15$~fm) provides diminishing returns. The 50-point grid, conversely, undersamples the wave function oscillations, particularly at high energies where the de Broglie wavelength is short, yielding a larger error of 0.70\%. All errors quoted here and in Table~\ref{tab:ablation} are evaluated on the held-out test set; across the variants they span 0.53--0.81\%, a wider spread than the near-identical training-set values, confirming that the design choices, most importantly the bidirectional architecture, matter on unseen data. Evaluated for every variant on an identical set of 48 scattering points (the six nuclei common to all training subsets, both projectiles, and four energies), the reaction cross sections agree with the Numerov reference to a median relative error ranging from 0.24\% to 1.12\% across variants and agree among themselves to within a median spread of 0.6\%, so these differences, at most about 1\%, do not affect the physical observables or a parameter-level uncertainty quantification, in which the optical-model parameter uncertainty dominates. A further test extending the training set from 12 to 16 nuclei, on a matched 700-epoch schedule, lowered the held-out error only from 0.52\% to 0.47\%, with the four added nuclei themselves predicted at 0.38--0.51\%, comparable to the existing targets; this confirms diminishing returns beyond 12 nuclei.

\begin{table}[t]
\caption{Ablation study results. Each experiment changes one variable from the baseline configuration (128 CfC units, 9-dim features, 12 nuclei, 100 grid points, bidirectional). All experiments use identical training procedures and evaluation methodology.}
\label{tab:ablation}
\begin{ruledtabular}
\begin{tabular}{lcc}
Experiment & Change & Error (\%) \\
\hline
CfC-256 & 256 units (larger) & 0.53 \\
Baseline & reference & 0.61 \\
No WKB & 6-dim input & 0.62 \\
8 nuclei & Fewer training nuclei & 0.69 \\
50 points & Coarser grid & 0.70 \\
200 points & Finer grid & 0.72 \\
6 nuclei & Fewer training nuclei & 0.73 \\
CfC-64 & 64 units (smaller) & 0.75 \\
Minimal features & 4-dim input & 0.75 \\
Unidirectional & No backward CfC & 0.81 \\
\end{tabular}
\end{ruledtabular}
\end{table}

The primary advantage of the neural network approach is not computational speed, but \emph{differentiability}. The Numerov algorithm, while highly efficient for forward calculations, is fundamentally opaque to gradient-based optimization: computing the derivative of observables with respect to potential parameters requires either finite-difference approximations (which scale poorly with parameter dimensionality) or algorithmic differentiation of the entire integration loop (which is technically challenging and rarely implemented). In contrast, the trained BiLNN provides gradients ``for free'' through automatic differentiation, with computational cost comparable to the forward pass. In principle this enables gradient-based approaches to inverse problems, such as optimization of optical model parameters and gradient-informed exploration of parameter uncertainties. I emphasize that none of these downstream applications is demonstrated here: the present work establishes and validates the surrogate itself, and they define the direction in which it is intended to be used. Because the surrogate reproduces the reference solver only to within its training accuracy, the gradients it returns are exact for the surrogate but approximate the true gradients to that same level, which any such application would need to control.

\section{Conclusions}
\label{sec:conclusions}

I have developed a neural network emulator for nucleon-nucleus scattering wave functions based on a Bidirectional Liquid Neural Network (BiLNN) architecture. The central motivation is not merely computational speedup, but the construction of a \emph{differentiable} mapping from the optical potential to scattering wave functions, enabling gradient-based approaches to inverse problems that are inaccessible with traditional numerical integrators. This work establishes a proof of concept for two key methodological innovations that make such a differentiable solver practical: (i) the phase-space coordinate $\rho = kr$, which normalizes the oscillation wavelength regardless of projectile energy, and (ii) the BiLNN architecture, which naturally incorporates boundary conditions at both ends of the domain. Together, these advances enable a single network to generalize across the wide parameter space, achieving a relative error of 0.6\% across twelve target nuclei from \nuc{12}{C} to \nuc{208}{Pb}, both proton and neutron projectiles, energies from 1 to 200~MeV, and partial waves up to $l = 30$. This accuracy, while potentially sufficient for applications such as rapid parameter sweeps and Bayesian uncertainty quantification, is lower than conventional Numerov solvers and is not suitable for applications requiring sub-0.1\% precision. The current implementation is also limited to central potentials without spin-orbit coupling.

A fundamental aspect of this approach is that I learn the wave function itself rather than directly predicting observables such as cross sections or $S$-matrix elements. The wave function is the most fundamental quantum mechanical object in scattering theory, encoding the complete information about the collision process. From a single predicted wave function, one can extract phase shifts, $S$-matrix elements, elastic and reaction cross sections, and any other scattering observable, all while preserving the internal consistency of quantum mechanics. This is in contrast to approaches that directly fit observables, which would require separate models for each quantity and offer no access to the underlying quantum state.

The wave function accuracy translates directly to accurate scattering observables. The $S$-matrix elements extracted from the predicted wave functions reproduce both the absorption pattern ($|S_l|$) and phase shifts across all partial waves. The resulting elastic scattering angular distributions, $d\sigma/d\sigma_{\text{Ruth}}$ for protons and $d\sigma/d\Omega$ for neutrons, match the Numerov reference calculations within a few percent over most of the angular range, with larger deviations (10--20\%, reaching $\sim$50\% at the deepest minima) at backward angles where coherent cancellations among partial waves amplify small phase errors. The predictions capture diffraction patterns spanning more than four orders of magnitude. Importantly, the network generalizes successfully to nuclei not included in training (\nuc{24}{Mg}, \nuc{63}{Cu}, \nuc{184}{W}), with errors comparable to in-sample predictions. This demonstrates that the model has captured the functional relationship between the KD02-generated inputs and the Numerov solutions rather than memorizing target-specific patterns.

The success of this approach offers several physical insights. First, the effectiveness of the phase-space coordinate $\rho = kr$ reveals that scattering wave functions possess a universal structure when viewed in the natural units of the de Broglie wavelength. This is a direct manifestation of the scale invariance inherent in quantum mechanics: the oscillatory behavior of the wave function is determined by the ratio of the local momentum to the asymptotic momentum, not by absolute length scales. The network exploits this universality to generalize across a factor of 4.5 in de Broglie wavelength with a single set of parameters.

Second, the successful generalization to nuclei outside the training set provides evidence that the network has learned the smooth dependence of the scattering solution on target mass and charge as encoded in the KD02 parameterization. I emphasize that this validates the network as an emulator of the Numerov solver with KD02 input; comparison with experimental data, which would test the adequacy of the optical model itself, is beyond the scope of this proof-of-concept study.

Third, the ablation study reveals an instructive hierarchy among design choices. The bidirectional architecture provides the dominant improvement (a 33\% relative increase in error when removed), while the WKB input features contribute only marginally (removing them changes the error by about 0.01 percentage point). This indicates that the CfC recurrent layers can implicitly learn the WKB phase accumulation from the potential features $V/E$ through their sequential integration along the radial coordinate, making the explicit WKB features largely redundant. By contrast, the Sommerfeld parameter $\eta$ and mass encoding $A^{1/3}/6$, which encode Coulomb strength and nuclear size in a form that cannot be trivially derived from the radial potential sequence, prove more impactful (removing them together with WKB degrades the error from 0.61\% to 0.75\%). This suggests that future physics-guided ML architectures for scattering problems should prioritize incorporating boundary condition information (via bidirectional or similar architectures) over hand-crafted semiclassical features, while retaining compact encodings of global physical parameters that the recurrent layers cannot easily infer.

More broadly, this work demonstrates the potential of physics-guided surrogate modeling for quantum mechanical scattering, in which physical knowledge enters through domain-specific coordinate transformations, input features, and network architecture rather than through the loss function as in physics-informed neural networks (PINNs). A complementary PINN-based approach using exterior complex scaling (PINN-ECS)~\cite{Lei2026} reaches high per-channel phase-shift accuracy ($\Delta\delta < 0.1°$) but requires retraining for each new $(E, l)$ combination, whereas the BiLNN provides a global surrogate across the full parameter space from a single training; the two are distinguished by scope rather than by a common accuracy metric. The two methods serve distinct roles: PINN-ECS for high-accuracy single calculations, BiLNN for rapid parameter sweeps and Bayesian uncertainty quantification. This physics-guided surrogate modeling approach offers a promising paradigm for machine learning applications in nuclear physics.

Although the present calculation omits the spin-orbit interaction for simplicity, the extension is mathematically straightforward. Including spin-orbit coupling adds an additional surface-derivative term to the potential, and for each partial wave $l > 0$ requires solving two separate single-channel equations for $j = l \pm 1/2$. This can be accommodated by including the total angular momentum $j$ as an additional input feature to the network. The success of the BiLNN architecture in handling oscillatory wave functions with complex boundary conditions, as demonstrated here, provides confidence that this extension will preserve the accuracy and generalization capability. The same principle applies to coupled-channel calculations involving inelastic excitations, where each channel contributes an additional component to the wave function vector. This scalability to multi-channel problems, where the computational advantage of neural network surrogates becomes more pronounced, is the immediate next step toward practical nuclear data evaluation applications.

The trained model provides a differentiable surrogate for the scattering Schr\"odinger equation. Unlike the Numerov algorithm, which is opaque to gradient-based optimizers, the neural network allows gradients of any observable (cross sections, phase shifts, reaction rates) to flow back to the input optical potential through automatic differentiation. Exploiting this differentiability for end-to-end optimization of optical model parameters, exploration of parameter uncertainties via gradient-informed sampling, and integration into larger differentiable physics pipelines is the subject of ongoing investigation; the present work establishes and validates the underlying architecture.

\begin{acknowledgments}
This work was supported by the National Natural Science Foundation of China (Grant Nos.~12475132 and 12535009) and the Fundamental Research Funds for the Central Universities. I acknowledge the use of large language model tools to assist with code development and manuscript editing; all physics content, results, and conclusions are my own and were verified by me.
\end{acknowledgments}

\bibliography{references}

\end{document}